\documentclass[%
reprint,
nofootinbib,
amsmath,amssymb,
aps,
]{revtex4-2}

\usepackage{graphicx}
\usepackage{dcolumn}
\usepackage{bm}
\usepackage[normalem]{ulem}
\usepackage{xcolor}

\usepackage{multirow}

\newcommand{\rb}{\bar{r}}
\newcommand{\tl}{\tilde{\lambda}}
\newcommand{\cH}{\mathcal{H}}

\newcommand{\be}{\begin{equation}}
\newcommand{\ee}{\end{equation}}
\newcommand{\bi}{\begin{itemize}}
\newcommand{\ei}{\end{itemize}}

\begin{document}
	
	
	\title{On the impact of perturbative counterterms on black holes}
	
	\author{Jesse Daas}
	\email{J.Daas@science.ru.nl}
	\author{Cristobal Laporte}%
	\email{cristobal.laportemunoz@ru.nl}
	\author{Frank Saueressig}%
	\email{f.saueressig@science.ru.nl}
	\affiliation{%
		Institute for Mathematics, Astrophysics and Particle Physics (IMAPP) \\ Radboud University, Heyendaalseweg 135, 6525 AJ Nijmegen,The Netherlands 
	}%

	\date{\today}
	
	\begin{abstract}
We determine the corrections to the Schwarzschild geometry arising from including the Goroff-Sagnotti counterterm in the gravitational dynamics. We find that static, asymptotically flat, and spherically symmetric geometries are completely characterized by their asymptotic mass and the coupling associated with the counterterm. The latter induces distinct corrections at sixth order of the parameterized post-Newtonian expansion. The resulting spacetime geometries still exhibit an event horizon. In the parameter space accessible to numerical integration, the horizon area is smaller than its Schwarzschild counterpart, leading to an increase in the Hawking temperature. Corrections to the shadow size can be determined analytically and are used to give a first bound on the new coupling. While it is difficult to access the geometry inside of the event horizon, our analysis also provides evidence that the counterterm could resolve the curvature singularity appearing in the Schwarzschild geometry.  		
	\end{abstract}
	
	\maketitle
	
\section{Introduction}
\label{sec.intro}
The detection of gravitational waves from binary black hole mergers \cite{LIGOScientific:2016aoc,KAGRA:2021duu} and shadow imaging of supermassive black holes \cite{EventHorizonTelescope:2019dse,EventHorizonTelescope:2019ggy,EventHorizonTelescope:2022wkp} have put a spotlight on black hole research. Most of the physics insights gained in this way rely on simulations which model the black hole spacetimes by the vacuum solutions of classical general relativity, i.e., the Schwarzschild spacetime and its rotating extension given by the Kerr solution. While this is perfectly reasonable at the classical level, the tension between general relativity and quantum mechanics is a strong pointer that this is not the final picture. On general grounds any theory of quantum gravity is expected to modify the classical black hole spacetimes in one way or another, e.g., when aiming at resolving the spacetime singularities featured by the classical solutions \cite{Bambi:2023try} or when dealing with the information paradox \cite{Chen:2014jwq}.

This letter takes an important step towards understanding black holes beyond general relativity. Concretely, we determine the corrections to the Schwarzschild geometry arising from the two-loop counterterm found by Goroff and Sagnotti \cite{Goroff:1985sz,Goroff:1985th,vandeVen:1991gw}. We show that the admissible spherically symmetric, asymptotically flat and static geometries have ``no hair'' in the sense that they are completely characterized by their asymptotic mass and the value of the new coupling. Moreover, we show that the event horizon persists and that the solution in the exterior can be accurately approximated by a simple analytic form. We also argue for singularity resolution, provided that the external solution connects to our local analysis at the core. A more detailed analysis will be provided in \cite{GS-inprep}.

\section{General Motivation and Setup}
\label{sec.2}
Constructing alternatives to the black hole spacetimes known from general relativity has, by now, evolved into a rich subject \cite{Lu:1993sq,Bueno:2016xff,Hennigar:2016gkm,Bueno:2016lrh,Svarc:2020fia,Knorr:2022kqp,Bueno:2022res,Bonanno:2022ibv,Silveravalle:2022wij,Held:2021pht,Aguilar-Gutierrez:2023kfn,Pawlowski:2023dda,Alvarez:2022lnf}. From a phenomenological perspective, one may construct models geared towards resolving deficits intrinsic to the black holes of general relativity. This motivates, e.g., the study of regular black hole spacetimes \cite{Bardeen:1968,Dymnikova:1992ux,Bonanno:2000ep,Frolov:2016pav} with the Hayward solution \cite{Hayward:2005gi} being one of the most prominent examples. Alternatively, one may study black holes based on classical extensions of general relativity. Prototypical examples are the black holes arising in scalar-tensor theories \cite{Sotiriou:2011dz,Hui:2012qt}, recently reviewed in \cite{Herdeiro:2015waa}, where the scalar field is used to model a generic fifth force. 

From the perspective of a quantum theory, one can regard the Einstein-Hilbert action as the leading term in an effective field theory \cite{Donoghue:2017pgk}. The gravitational dynamics  receives corrections in the form of local higher-derivative terms and non-local contributions encoded in form factors. The effect of the latter on the Schwarzschild geometry has been considered in \cite{Calmet:2021stu,Koshelev:2018hpt,Buoninfante:2018rlq,Boos:2021kqe}. Restricting to the local part, one-loop corrections lead to new terms in the dynamics which are quadratic in the spacetime curvature and one naturally arrives at quadratic gravity \cite{Stelle:1976gc,Stelle:1977ry}. These contributions vanish when the vacuum Einstein equations are imposed. From the perspective of quantizing general relativity, this is the statement that the one-loop counterterms are either topological or can be absorbed by a field redefinition \cite{tHooft:1974toh}. For the spacetime geometry, this entails that solutions of classical general relativity are also solutions of quadratic gravity. While the solution space of the latter is significantly richer \cite{Lu:2015tle,Lu:2015cqa,Lu:2015psa,Pravda:2016fue,Svarc:2018coe,Podolsky:2018pfe,Podolsky:2019gro,Pravda:2020zno,Bonanno:2019rsq,Bonanno:2021zoy,Daas:2022iid,Silveravalle:2022lid},
any confirmation of vacuum properties of general relativity can also be interpreted as a confirmation of quadratic gravity \cite{Daas:2022iid}.

In order to arrive at a non-vanishing correction to the Schwarzschild geometry, one needs to go beyond the terms quadratic in the spacetime curvature. Taking inspiration from the perturbative quantization of general relativity, one may then ask what is the ``simplest'' local modification, coming with the lowest number of spacetime derivatives, which expels the Schwarzschild geometry as a solution of the equations of motion. The unique answer to this question is the Goroff-Sagnotti counterterm, coming with six spacetime derivatives. This correction exhibits several outstanding properties. Firstly, its contribution to the gravitational dynamics does not vanish when the vacuum Einstein equations are imposed. Hence it gives rise to genuine modifications of the black hole solutions known from general relativity. Secondly, since one is dealing with a counterterm in the quantization of general relativity, it is expected to appear in the effective gravitational dynamics in a rather universal way. Thirdly, we verified that the counterterm does not give rise to additional degrees of freedom, at least as a perturbation around a flat background. Thus the Yukawa-type interactions associated with the massive ghost-modes appearing in quadratic gravity are absent in this case.  These properties provide a strong motivation for studying static, spherically symmetric spacetimes with this particular contribution added to the dynamics. 

Following the path motivated above, our starting point is the Einstein-Hilbert (EH) action with the cosmological constant set to zero,
\be\label{EHaction}
S^{\rm EH}[g] = \frac{1}{16 \pi G} \int d^4x \sqrt{-g} R \, , 
\ee
where $G$ is Newton's constant and $g \equiv \det(g_{\mu\nu})$ is the determinant of the spacetime metric $g_{\mu\nu}$. The covariant derivative, Ricci scalar, Ricci tensor and Weyl tensor constructed from $g_{\mu\nu}$ are denoted by $D_\mu$, $R$, $R_{\mu\nu}$ and $C_{\mu\nu\rho\sigma}$, respectively. The dynamics \eqref{EHaction} is complemented by the Goroff-Sagnotti (GS) term,
\be\label{GSaction}
S^{\rm GS}[g] = \frac{G^2 \, \lambda}{16 \pi} \int d^4x \sqrt{-g} \left( C_{\mu\nu}^{\phantom{\mu\nu}\kappa\gamma} \, C_{\kappa\gamma}^{\phantom{\kappa\gamma}\rho\sigma} \, C_{\rho\sigma}^{\phantom{\rho\sigma}\mu\nu}  \right) \, , 
\ee
where $\lambda$ is an a priori undetermined, dimensionless coupling constant. For later purposes, it is convenient to introduce $\tl \equiv \lambda G^2$ as a short-hand. The perturbative quantization of \eqref{EHaction} necessitates the inclusion of \eqref{GSaction} in order to absorb the infinities appearing in the quantization procedure at two loop-level \cite{Goroff:1985sz,Goroff:1985th,vandeVen:1991gw}. 

The equations of motion are obtained from the variation principle
\be\label{eom}
H_{\mu\nu} \equiv \frac{1}{\sqrt{-g}} \, \frac{\delta}{\delta g^{\mu\nu}} \left( S^{\rm EH} + S^{\rm GS} \right) = 0 \, . 
\ee		
The explicit form of $H_{\mu\nu}$ is readily obtained using computer algebra packages like the xAct package for {\tt Mathematica} \cite{xactref}. The result is lengthy and little illuminating, so we refrain from giving it explicitly. The fact that $H_{\mu\nu}$ is obtained from a diffeomorphism invariant action functional entails that the components appearing in \eqref{eom} are related by $D^\mu H_{\mu\nu} = 0$.

We are interested in static, spherically symmetric solutions of the system \eqref{eom}. Using Schwarzschild coordinates, the most general form of the line-element compatible with these symmetry requirements is \cite{Hawking:1973uf}
\be\label{SS:coords}
ds^2 = - h(\rb) dt^2 + f(\rb)^{-1} d\rb^2 + \rb^2 \left(d\theta^2 + \sin^2 \theta d\phi^2 \right) \, . 
\ee		
where the radial coordinate $\rb$ is defined geometrically via the area of the two-spheres coordinatized by $\theta, \phi$. For the Schwarzschild geometry, the two metric functions $h(\rb)$ and $f(\rb)$ agree and are given by
\be\label{SS:geometry}
h(\rb) = f(\rb) = 1 - \frac{2 G M}{\rb} \, . 
\ee
Here $M$ denotes the asymptotic mass of the geometry. Due to the contributions of the Goroff-Sagnotti term in \eqref{eom}, the geometry \eqref{SS:geometry} is not a solution of the modified equations of motion though.		

In order to study the solutions of the system \eqref{eom} efficiently, we follow a strategy highly successful in the context of quadratic gravity \cite{Pravda:2016fue,Svarc:2018coe,Podolsky:2018pfe,Podolsky:2019gro,Pravda:2020zno}, and recast the line-element \eqref{SS:coords} into conformal-to-Kundt form
\be\label{CtK:coords}
ds^2 = \Omega^2(r) \left( d\theta^2 + \sin^2\theta d\phi^2 - 2 du dr + \cH(r) du^2 \right) \, .
\ee
The coordinate transformation relating \eqref{SS:coords} and \eqref{CtK:coords} is
$
\bar{r} = \Omega(r), t = u - \int \frac{dr}{\mathcal{H}(r)}$
and the metric functions obey
\be\label{rel.metfcts}
 h = - \Omega^2(r) \, \cH(r) \, , \quad f = - \left(\frac{\Omega'(r)}{\Omega(r)}\right)^2 \, \cH(r) \, .
\ee
A direct consequence of these relations is that the center of the geometry, $\rb = 0$, is mapped to $\Omega(r) = 0$. Horizons, satisfying $f(\rb_h) = h(\rb_h) = 0$, are encoded in the roots $\cH(r_h) = 0$ with $\Omega(r_h) = \rb_h$ being finite.

The main advantage of the conformal-to-Kundt description is that it does not contain any explicit $r$-dependence. As a consequence, the equations of motion for the metric functions $\Omega(r)$ and $\cH(r)$, resulting from substituting the line-element \eqref{CtK:coords} into \eqref{eom}, are an \emph{autonomous} system of coupled, non-linear differential equations $H_{rr} = H_{ru} = H_{\theta\theta} = 0$ which contain fourth derivatives of $\Omega$ and $\cH$ with respect to $r$. The equations $H_{ru} = 0$ and $H_{\theta\theta} = 0$ can be written as
\be\label{eomsys}
\begin{split}
\cH^{\prime\prime \prime} & =  \tfrac{ \tl \left(3 \cH^\prime \Omega^\prime + \Omega (\cH^{\prime\prime}-1)\right) \left(2 + \cH^{\prime\prime}\right)^2 - 54 \cH \Omega^3 (\Omega^{\prime})^2 - 18 \cH^\prime \Omega^\prime \Omega^4 - 18 \Omega^5}{3 \, \tl \, \Omega \, \cH^\prime \left(2 + \cH^{\prime\prime}\right)},  
\\  
\Omega^{\prime\prime} & =  \tfrac{\tl \left(2 + \cH^{\prime\prime}\right)^3 - 18 \Omega^4 (2+ \cH^{\prime\prime}) - 108 \Omega^3 \Omega^\prime \cH^\prime}{108 \cH \Omega^3} \, . 
\end{split}
\ee
Here the primes denote derivatives with respect to $r$ and we solved for the highest derivative terms. Once a solution of this system is found one checks that it also satisfies $H_{rr} = 0$.
%

\section{Black hole spacetimes}
\label{sec.3a}
While the conformal-to-Kundt prescription of the geometry leads to a significant simplification of the equations of motion, an analytic solution is currently still out of reach. A first step towards understanding the impact of the correction term then consists in determining local solutions of the system, which can be obtained as a series expansion at an arbitrary point $r_0$ through the Frobenius method. Building on the conformal-to-Kundt coordinates, the generic form of this expansion is\footnote{Our analysis covers solutions which admit a series expansion at the expansion point only. Tracking non-analytic terms is beyond the present work.}
\be\label{ctk:expansion}
\begin{split}
\Omega(r) = & (r-r_0)^n \sum_{i=0} a_i \, (r-r_0)^i \, , \\
\cH(r) = & (r-r_0)^p \sum_{i=0} b_i \, (r-r_0)^i  \, .
\end{split}
\ee
Here $(n,p)$ are the scaling exponents of the solution and we have $a_0 \not = 0$ and $b_0 \not = 0$ by definition. 

Substituting these expansions into the system \eqref{eomsys} and matching terms with equal powers of $(r-r_0)^i$ leads to a hierarchy of equations which determines the admissible values for $(n,p)$ as solutions of the incidental polynomial. The free coefficients $a_i, b_i$ parameterizing the solution space typically appear at low index numbers and the hierarchy of equations then expresses the higher-order expansion parameters in terms of these free parameters.

\subsection{Corrections in the asymptotically flat region}
\label{sec.31}
We start by studying the corrections in the asymptotically flat regions. For this purpose, we use the Schwarzschild coordinates and expand the metric functions in terms of inverse powers of $\rb$:\footnote{In the case of quadratic gravity, this expansion fails to capture the contributions of the massive graviton excitations which are non-analytic in the asymptotically flat region \cite{Saueressig:2021wam}. In the present framework, we find these modes are absent. Thus a polynomial expansion is expected to work.}
\be\label{asymansatz}
h(\rb) = \sum_{n=0} h_n \, \rb^{-n} \, , \quad f(\rb) = \sum_{n=0} f_n \, \rb^{-n} \, , 
\ee
with $h_0 = f_0 = 1$. Substituting this expansion into the corresponding equations of motion and comparing powers of $1/\rb$ yields
\be\label{asmptcorrections}
\begin{split}
h(\rb) = & 1 -\frac{\rb_h}{\rb} + \frac{6 \, \tl \, \rb_h^2}{\rb^6} - \frac{4 \, \tl \, \rb_h^3}{\rb^7} + O\left(\frac{\tl^2\, \rb_h^3}{\rb^{11}} \right) \, ,  \\
f(\rb) = & 1 - \frac{\rb_h}{\rb} + \frac{18 \, \tl \, \rb_h^2}{\rb^6} - \frac{16\,\tl \, \rb_h^3}{\rb^7} + O\left(\frac{\tl^2\, \rb_h^3}{\rb^{11}} \right) \, ,  
\end{split}
\ee
where we abbreviated $\rb_h \equiv 2 G M$. For $\tl = 0$ this expression reduces to the Schwarzschild geometry \eqref{SS:geometry} and we verified that the expansion using conformal-to-Kundt coordinates gives the same result. 

Eq.\ \eqref{asmptcorrections} is remarkable for three reasons. Firstly, the solution is determined completely in terms of the asymptotic mass $M$ and the new coupling $\tl$. Thus the solution exhibits a ``no-hair''-type property in the sense that it is fixed completely once the asymptotic mass and the coupling $\tl$ are given. Secondly, the leading corrections occur at $O(\rb^{-6})$ only \cite{Anselmi:2013wha,deRham:2020ejn}. The numerical coefficients at order $\tl$ agree with the published version of \cite{Alvarez:2023gfg}. Converting the line-element to isotropic coordinates shows that this corresponds to corrections at the sixth order in the parameterized post-Newtonian (PPN) expansion. The lower-order PPN parameters agree with general relativity. Power-counting arguments suggest corrections at fifth order in the PPN expansion already. Since the (leading) contributions of the GS-term are of second order in the spacetime curvature, $\tl$ appears in the combination $\tl M^2$ only. Hence the fifth order corrections are absent. Thirdly, the corrections break the degeneracy in the metric functions so that for $\tl \not = 0$ one has $h(\rb) \not = f(\rb)$ with \emph{a fixed ratio between the leading correction terms}. This characteristic fingerprint invites probing \eqref{asmptcorrections} with experiments that are sensitive to a difference in the metric functions. 

\subsection{Admissible local solutions}
\label{sec.32}
We proceed by analyzing the local solutions arising from \eqref{eomsys} using a Frobenius analysis based the ansatz \eqref{ctk:expansion}.
%
 The independent solution classes identified in this way are summarized in Table \ref{Tab.1} and we highlight their most important properties.
\begin{table}[t]
\renewcommand{\arraystretch}{1.3}
\begin{tabular}{rcc}
	$(n,p)^{\phantom{\infty}}$ & expansion point & free parameters \\ \hline \hline
	$(0,0)^{\phantom{\infty}}$ $\Bigg.$ & regular point & \quad $a_0, a_1, b_0, b_1, b_2; r_0$ \quad \\ \hline
\multirow{3}*{$(-1,2)^0_{\phantom{\infty}}$}	 & \quad  asymptotically flat region \quad  & $a_0, b_0, b_1=-2M$  \\ 
 & $(r \rightarrow 0, r_0 = 0)$ &  $a_0, b_0$ fixed by \\
 && asymptotic flatness  \\[1.2ex] \hline
 \multirow{2}*{$(-1,2)^\infty$}	 & center of the geometry &  \\
 & $(r \rightarrow \infty, r_0 = 0)$ &  \\[1.2ex] \hline
	$(0,1)^{\phantom{\infty}}$ $\Bigg.$ & horizon &$a_0, b_0, b_1; r_0 = r_h$  \\ 
\hline
$(0,2)^{\phantom{\infty}}$ $\Bigg.$ & double horizon &$a_0; r_0 = r_h$ \\
\hline \hline
\end{tabular}
\caption{\label{Tab.1} Classification of the admissible local scaling behaviors arising from solving the system \eqref{eomsys} based on the ansatz \eqref{ctk:expansion}.}
\end{table}

 The \emph{asymptotically flat region} is covered by the solutions $(-1,2)^0$ coming with three free parameters. The condition that the geometry approaches Minkowski space fixes $a_0$ and $b_0$. This is the no-hair property highlighted in eq.\ \eqref{asmptcorrections}. At a \emph{generic point}, both $\Omega(r)$ and $\cH(r)$ are finite by definition. This is reflected in the $(0,0)$ class. Keeping $r_0$ fixed, this class has five free parameters. Two of these may be fixed by a canonical normalization of $t,r$ in the asymptotically flat region. The fact that one is then left with more free parameters than in the $(-1,2)^0$ case is not a contradiction per se: not all solutions found in the interior need to connect to the asymptotically flat region. Our analysis also shows that there is a consistent family of solutions $(0,1)$. This ensures that the equations of motion actually admit geometries exhibiting an event horizon. The horizon constitutes a singular point of the system \eqref{eomsys} though. As a consequence, a generic solution may not admit a continuation into the interior of the black hole.

The most intriguing results are obtained in the core region described by the $(-1,2)^\infty$ solution. In this case the analysis is very involved since one has to track a significant number of branches due to the non-linear appearance of the coefficients $a_i, b_i$ in the hierarchy. Our analysis suggests that there is one non-trivial, consistent solution. This solution has the remarkable property that all sixteen Carminati–McLenaghan invariants \cite{Carminati:1991ddy}, including the Kretschmann scalar $R_{\mu\nu\rho\sigma} R^{\mu\nu\rho\sigma}$ and the square of the Weyl tensor, remain finite as $\rb \rightarrow 0$. This is in line with earlier arguments \cite{Holdom:2002xy}. Thus the class does not admit spacetime singularities. This is in stark contrast to a similar analysis in quadratic gravity where the higher-derivative terms lead to an amplification of the curvature singularity for asymptotically flat solutions \cite{Podolsky:2019gro,Daas:2022iid}.

\begin{figure}[t]
	\centering
	\includegraphics{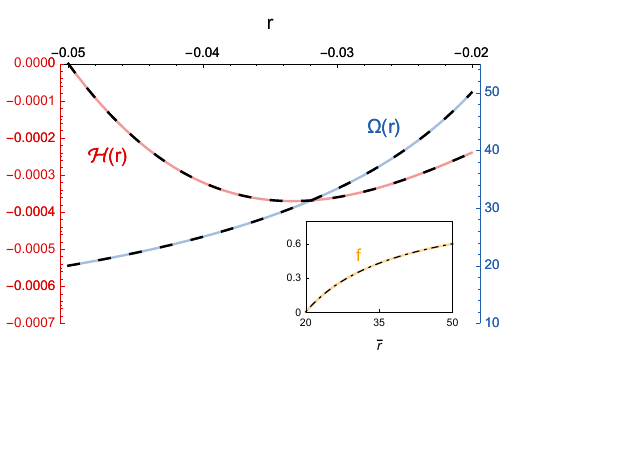}
	\vspace{-26mm}
	\caption{Black hole solution obtained from integrating the system \eqref{eomsys} numerically with initial conditions $M=10$, $\lambda = 0.1$ set at $r_{\rm init} = -0.02$. This solution is representative for geometries obtained for $\lambda = [0, c M^4]$ where $c$ is a positive constant of order unity. This class is characterized by, firstly, the existence of an event horizon and, secondly, following the asymptotics \eqref{asmptcorrections} in the asymptotically flat region. The main diagram shows the metric functions $\mathcal{H}(r) < 0$ (left-side scale) and $\Omega(r) > 0$ (right-side scale) describing the black hole in the conformal-to-Kundt coordinates. Its horizon is located at $r_h^{\rm GS} = -0.0500000423$. The solution closely follows the Schwarzschild solution (superimposed as black dashed lines) up to the horizon. Inset: comparison between the metric function $f(\rb)$ of the numerical solution (orange line) and the approximation \eqref{asmptcorrections}, superimposed as a dashed-dotted line. The difference between the exact solution and the analytic approximation is found to be of the order $10^{-6}$. The metric function $h(\rb)$ exhibits the same features. }
	\label{fig:enter-label}
\end{figure}
%
\subsection{Numerical integration}
\label{sec.3}
The Frobenius method gives local solutions at the core, the horizon, and the asymptotically flat region. In order to arrive at the global picture, we then integrate the system \eqref{eomsys} numerically, placing initial conditions in the asymptotically flat region.
Our analysis focuses on the case $\lambda > 0$. While $\lambda < 0$ can be plausible as well, the numerical integration of the equations of motion is much more involved in this case and we have not been able to connect the local solutions in this way.

A prototypical solution is shown in Fig.\ \ref{fig:enter-label}. Starting from the asymptotically flat region, the numerical integration of \eqref{eomsys}, reveals that the geometries exhibit an event horizon where $\cH(r)|_{r = r^{\rm GS}_h} = 0$. Moreover, the comparison between the numerical integration of \eqref{eomsys} and the analytic approximation \eqref{asmptcorrections} shows that the latter is reliable up to the horizon, at least in the parametric regime where corrections of the order $\tl^2/M^8$ are negligible. Thus the asymptotically flat solutions connect continuously to the local solutions at the event horizon and we obtain a comprehensive picture of the geometry in the outside region of the black hole.
%

\section{Thermodynamics and shadow imaging}
\label{sec.4}
The series expansion \eqref{asmptcorrections} gives a good description of the geometry in the outer region of the black hole. Thus we take it as a starting point for determining the position of the event horizon and its properties. All analytic results have been corroborated by integrating the exact equations numerically. \\

\noindent
\emph{Horizons and black hole thermodynamics.} Solving the horizon condition $f(\rb_h^{\rm GS};\lambda) = h(\rb_h^{\rm GS};\lambda) = 0$ perturbatively in $\lambda$ yields
\be\label{eq:rgs}
\rb_h^{\rm GS}(\tl) - \rb_h = - \frac{\tl}{4 \, G^3 M^3} + O\left( \tl^2 \right) \,  . 
\ee
Thus, for $\tl > 0$, the corrections due to the Goroff-Sagnotti term lead to black holes which have a smaller horizon area than their Schwarzschild counterparts with the same asymptotic mass $M$. 

\begin{figure}[h]
\includegraphics[width=0.95 \columnwidth]{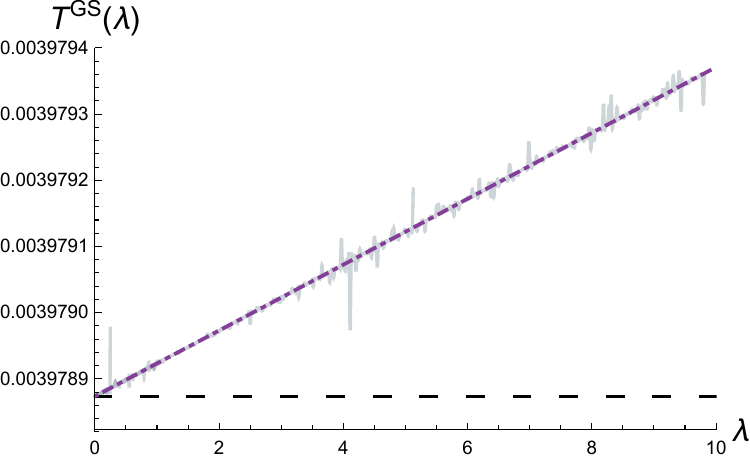}
\caption{\label{Fig.2} Dependence of the horizon temperature \eqref{eq:Tgs} on $\lambda$ for an asymptotic mass $M = 10$ ($G=1$). The dashed horizontal line gives the Schwarzschild result. The dashed-dotted line is the result \eqref{eq:Tgs} and we have superimposed the result from integrating the equations of motion numerically as the gray line. The spikes constitute an artifact of our numerical integration procedure.}
\end{figure}

 The horizon also allows to associate a horizon temperature,
%
$T \equiv \frac{\kappa}{2\pi}$. 
%
The surface gravity $\kappa$ is defined via $\xi^\mu D_\mu \xi^\nu = \kappa \xi^\nu$ and $\xi^\mu = (1,0,0,0)$ is the Killing vector associated with time translations. A brief computation based on the expansion \eqref{asmptcorrections} yields
\be\label{eq:Tgs}
T^{\rm GS}(\tl) - T^{\rm SS} = \frac{\tl}{16 \pi  \, G^5 M^5} + O\left( \tl^2 \right)  \, . 
\ee
Thus, $\tl > 0$ leads to an increase in the horizon temperature. This is illustrated in Fig.\ \ref{Fig.2} where the analytic result \eqref{eq:Tgs} is also confirmed by numerical integration.\\

\noindent
\emph{Bounding $\lambda$ from shadow observations.} A powerful tool for probing the spacetime geometry in the vicinity of a black hole horizon is taking an image resolving the central brightness depleted region (shadow imaging). The size of the shadow is determined by the photon ring located at coordinate radius $\rb_{\rm ph}$ where light rays can undergo an unstable circular orbit. For the geometry \eqref{SS:coords} one has \cite{EventHorizonTelescope:2020qrl,Daas:2022iid}
\be\label{eq:shadow}
\rb_{\rm ph} = 2 \, h(\rb_{\rm ph}) \left( \left. \frac{d h(\rb)}{d\rb} \right|_{\rb = \rb_{\rm ph}} \right)^{-1} \, , \;
\rb_{\rm sh} = \frac{\rb_{\rm ph}}{\sqrt{h(\rb_{\rm ph})}} \, .
\ee
Here the first relation is an implicit equation for $\rb_{\rm ph}$. The second identity determines the shadow size $\rb_{\rm sh}$, taking into account that the bending of light close to the black hole causes the photon ring to appear bigger than it really is. The angular radius of the photon ring on the sky is then given by $\theta_{\rm sh} = \rb_{\rm sh}/D$ with $D$ the distance to the object. For the $M_{87}^*$ black hole, one of two Event Horizon Telescope candidates, observations give $\theta_{\rm sh} = 42 \pm 3$ $\mu$as \cite{EventHorizonTelescope:2019dse}. 

For the Schwarzschild case the relations \eqref{eq:shadow} evaluate to
$
\rb_{\rm ph}^{\rm SS} = 3M, \rb_{\rm sh}^{\rm SS} = 3 \sqrt{3} \, M
$.
We then determine the impact of $\lambda$ on the shadow size, based on eq.\ \eqref{asmptcorrections}. Defining
$
\zeta \equiv \frac{\rb_{\rm sh}^{\rm GS}(\tl) - \rb_{\rm sh}^{\rm SS}}{\rb_{\rm sh}^{\rm SS}}, 
$ 
one obtains
\be
\zeta = - \frac{320}{729}\,\frac{\tl}{\rb_h^4} + \mathcal{O}\big(\frac{\tl^2}{M^8}\big)\,.
\ee
The sign indicates that for $\tl > 0$ the shadow size is decreased. This is consistent with the black hole being more compact. The Schwarzschild radius of $M^*_{87}$ is estimated as $\rb_h \approx 2 \times 10^{13}$ m. Asking that the correction term changes the angular radius by at most 10$\%$ then bounds the dimensionless parameter $|\lambda| < 5.4 \times 10^{191}$.

\section{Summary and outlook}
\label{sec.conclusions}
The need for theoretically well-motivated black hole spacetimes going beyond the framework of classical general relativity can hardly be overstated. In this letter, we contributed to this theme by determining the imprints of the Goroff-Sagnotti counterterm on the Schwarzschild geometry. Eq.\ \eqref{asmptcorrections} shows that this leads to characteristic correction terms which are excluded by the uniqueness theorems of general relativity and modify the geometry at sixth order in the PPN. Finding conclusive evidence for these extra terms would be a major step towards testing fundamental ideas about quantum gravity, since any quantum theory of the gravitational interactions has to address the role of the perturbative singularity canceled by the Goroff-Sagnotti term at some point. This includes, in particular, the gravitational asymptotic safety program  \cite{Percacci:2017fkn,Reuter:2019byg,Saueressig:2023irs,Eichhorn:2022gku,Pawlowski:2023gym}, where the value of the new coupling can be predicted from first principle computations \cite{Gies:2016con}. It would be very exciting to explore to which extend gravitational wave signals emitted in a binary black hole mergers are sensitive to our modifications.

		
		\begin{acknowledgments}
			We thank A.\ Bonanno, A.\ Held, B.\ Knorr, J.\ Oliva, J.\ Podolsky, A.\ Ribes, S.\ Silveravalle, M.\ Wondrak and J.\ Zanelli for insightful discussions. The work of C.L.\ is supported by the scholarship Becas Chile ANID-PCHA/2020-72210073. This publication is part of the Dutch Black Hole Consortium with project number NWA.1292.19.202 of the research programme NWA which is (partly) financed by the Dutch Research Council (NWO).
		\end{acknowledgments}


%

	\end{document}